\documentclass[dvips]{revtex4}
\usepackage[dvips]{graphicx}
\usepackage{amsmath}
\usepackage{amssymb}

\def\slashchar#1{\setbox0=\hbox{$#1$}
   \dimen0=\wd0 \setbox1=\hbox{/} \dimen1=\wd1
   \ifdim\dimen0>\dimen1 \rlap{\hbox to \dimen0{\hfil/\hfil}} #1
   \else  \rlap{\hbox to \dimen1{\hfil$#1$\hfil}} / \fi}

\begin{document}

\title{Production of Two Pions Induced by Neutrinos}

\author{M. Valverde}
\affiliation{Departamento de F\'\i sica At\'omica, Molecular y Nuclear,
\\Universidad de Granada, E-18071 Granada, Spain.}
\author{J. Nieves}
\affiliation{Departamento de F\'\i sica At\'omica, Molecular y Nuclear,
\\Universidad de Granada, E-18071 Granada, Spain.}
\author{E. Hern\'andez} 
\affiliation{Grupo de F\'\i sica Nuclear,
Departamento de F\'\i sica Fundamental e IUFFyM,\\ Universidad de
Salamanca, E-37008 Salamanca, Spain.}  
\author{S.K. Singh} 
\affiliation{Department of Physics, Aligarh Muslim
University, Aligarh-202002, India.}  
\author{M.J. \surname{Vicente Vacas}} 
\affiliation{Departamento de
F\'\i sica Te\'orica and IFIC, Centro Mixto Universidad de
Valencia-CSIC\\ Institutos de Investigaci\'on de Paterna,
Aptdo. 22085, E-46071 Valencia, Spain.}

\today
\begin{abstract}
We study the threshold production of two pions induced by neutrinos in
nucleon targets. The contribution of nucleon pole, pion and contact terms
is calculated using a chiral Lagrangian. The contribution of the
Roper resonance, neglected in earlier studies, has also been taken
into account.
\end{abstract}

\maketitle

\section{Introduction}

A proper and precise understanding of the processes induced by
neutrino interactions is required in the analysis of neutrino
oscillation experiments. For instance, at intermediate energies, above
$0.5$ GeV, one pion production becomes relevant. Most of the
theoretical models for this reaction assume the dominance of
$\Delta(1232)$ resonance
mechanism\cite{AlvarezRuso:1997jr,AlvarezRuso:1998hi,Paschos:2003qr,Lalakulich:2005cs},
but others also include background
terms\cite{Sato:2003rq,Hernandez:2007qq,Hernandez:2007eb,Hernandez:2007kb}.
Above these energies new baryonic resonances can be excited, the first
of these resonances being the Roper $N^*(1440)$ which has a sizable
decay into a scalar pion pair and it is very wide. However, the
$\Delta$ does not couple to two pions in $s$-wave and thus it is not
relevant at energies where only slow pions are produced.

There exist very few attempts to measure the two pion production
induced by neutrinos and antineutrinos. Experiments done at
ANL\cite{Barish:1978pj,Day:1984nf} and BNL\cite{Kitagaki:1986ct}
investigated the two pion production processes, in order to test the
predictions of chiral symmetry.  Biswas {\it et
al.}\cite{Biswas:1978ey} used PCAC and current algebra methods to
calculate the threshold production of two pions. Adjei {\it et
al.}\cite{Adjei:1980nj} made specific predictions using an effective
Lagrangian incorporating chiral symmetry. However, these models did
not include any resonance production, as we do. Furthermore we use an
expansion of the chiral Lagrangian that includes terms up to ${\cal
O}(1/f_\pi^3)$, while Adjei {\it et al.} kept only terms up to ${\cal
O}(1/f_\pi^2)$. More detailed discussions can be found in
Ref.~\cite{Hernandez:2007ej}.

\section{Pion Production Model}

\begin{figure}[h]
\centering\includegraphics[width=\textwidth]{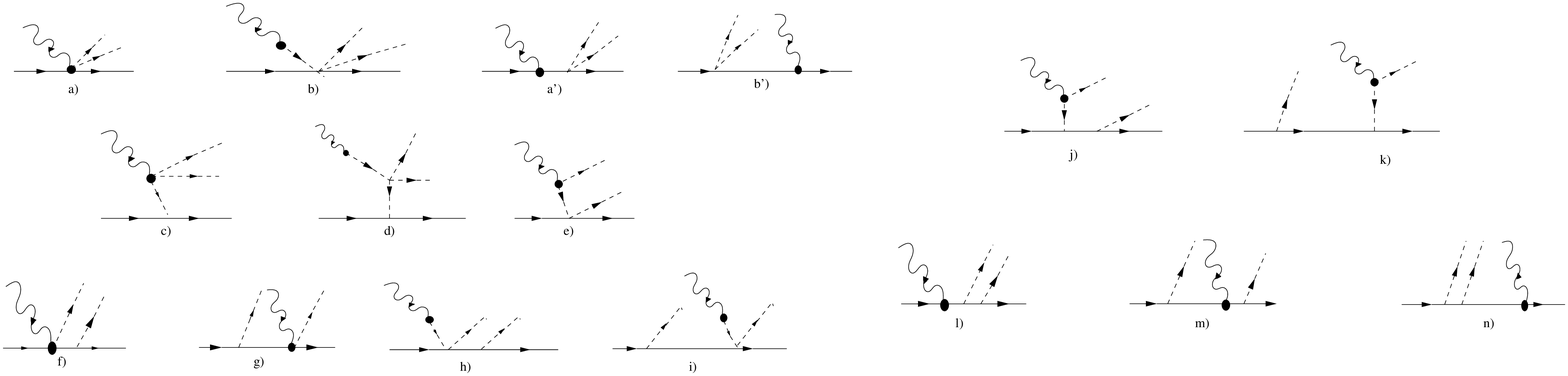}
\centering\includegraphics[width=\textwidth]{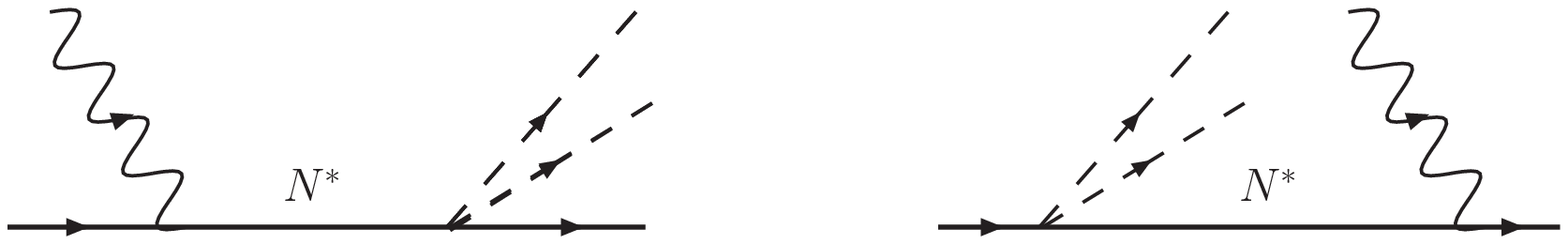}
\vspace*{8pt}
  \caption{Top: Nucleon pole, pion pole and contact terms contributing to
    $2\pi$ production.Bottom:  Direct and crossed Roper excitation
  contributions to $2\pi$ production. \protect\label{fig:fig1}}
\end{figure}

We will focus on the neutrino--pion production reaction off the
nucleon driven by charged currents,
\begin{equation}
\nu_l(k) + N(p) \to l^-(k^\prime) + N(p^\prime) + \pi(k_{\pi_1})+\pi(k_{\pi_2})
\label{eq:reac} \, .
\end{equation}
For the derivation of the hadronic current we use the effective
Lagrangian given by the SU(2) non-linear $\sigma$ model. This 
model\cite{Hernandez:2007qq} provides us with expressions for the
non-resonant hadronic currents that couple with the lepton current, in
terms of the first sixteen Feynman diagrams depicted in Fig.~\ref{fig:fig1}.

We also include the two mechanisms depicted in the bottom of
Fig.\ref{fig:fig1}, which account for the Roper production and its
decay into two pions in a $s$-wave isoscalar state.  The coupling of
the Roper to the charged weak current is written in terms of the
current
\begin{equation}
J^\alpha_{cc*} = 
\frac{F_1^{V*}(q^2)}{\mu^2}(q^\alpha\slashchar{q}-q^2\gamma^\alpha)
+ i\frac{F_2^{V*}(q^2)}{\mu}\sigma^{\alpha\nu}q_\nu 
- G_A\gamma^\alpha\gamma_5 - \frac{G_P}{\mu}q^\alpha\slashchar q\gamma_5 
- \frac{G_T}{\mu} \sigma^{\alpha\nu}q_\nu\gamma_5 \,,
\end{equation}
which is the most general form compatible with conservation of the
vector current. The $G_T$ term does not need to
vanish; however, most analyses neglect its contribution and we shall
do so here. The form factors $G_A$ and $G_P$ are constrained by PCAC
and the pion pole dominance assumption.  The vector form factors
$F_1^{V*}$ and $F_2^{V*}$ can be related to the isovector
part of the electromagnetic (EM) form factors.
We have fitted the proton-Roper EM transition form factors\cite{AlvarezRuso:2003gj} to the
experimental results for helicity amplitudes
\cite{Aznauryan:2004jd,Tiator:2003uu},
using a modiffied dipole parametrization (labeled FF1).  
The Roper EM data have large error bars and it is possible to
accommodate quite different functional forms and values for these FF.
Thus, we shall consider other different models for the vector form
factors: the constituent quark model of Meyer {\it et al.}\cite{Meyer:2001js} 
(FF2), the parametrizations of Lalakulich 
{\it et al.}\cite{Lalakulich:2006sw} (FF3) and finally the predictions of the
recent MAID\cite{Drechsel:2007if} analysis (FF4).

\section{Results}

\begin{figure}[h]
\centering\includegraphics[width=0.5\textwidth]{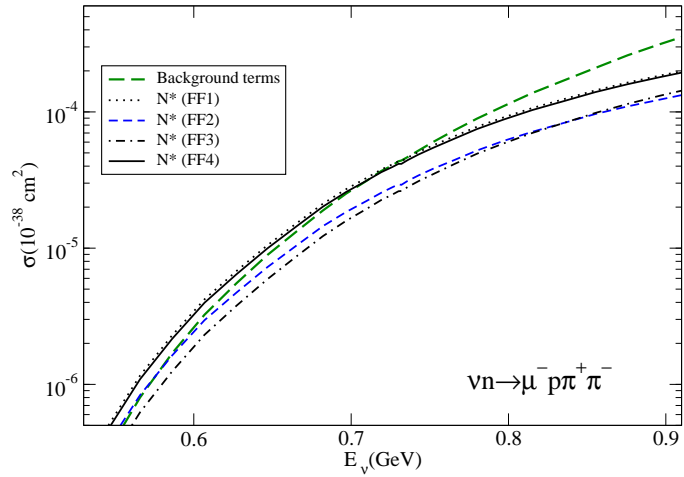}
\vspace*{8pt}
  \caption{Cross section for the $\nu n\to \mu^- p \pi^+\pi^-$
  reaction as a function of the neutrino energy. \protect\label{fig:3}}
\end{figure}

In Fig.~\ref{fig:3}, we present results for the cross section for
the process $\nu n\to \mu^- p \pi^+\pi^-$.  We show separately the
contribution of the background terms as well as the contribution of
the Roper resonance as calculated by using the various form factors
described above.  The interference between background and the Roper
contribution is not shown.  We see that the background terms dominate
the cross section for neutrino energies $E_\nu>0.7$ GeV. At lower
energies the contribution from the Roper could be larger or smaller
than the background depending upon the vector form factors used for
the $W^+NN^*$ transition. The differences in the predictions for the
cross sections using the various parametrizations could reach a factor
two.  The Roper contribution is specially sensitive to $F_2^{V*}(q^2)$
which is negative in contrast to the positive value which one gets in
the case of the nucleon.

\begin{figure}[h]
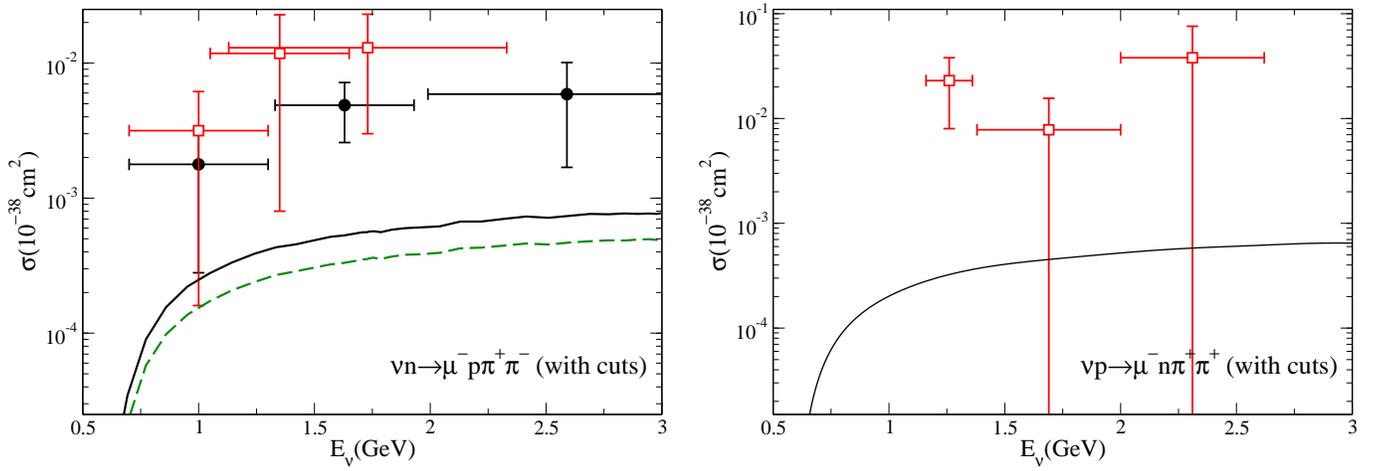

\centering{\mbox{\includegraphics[width=0.49\textwidth]{Fig5.FF1.eps}
                  \hspace{0.01\textwidth}
\includegraphics[width=0.49\textwidth]{Fig6.eps}}}
\vspace*{8pt}
\caption{Cross section for the $\nu n\to \mu^- p \pi^+\pi^-$ (left)
and $\nu p\to \mu^- n \pi^+\pi^+$ (rights) with cuts as explained in
the text.  Dashed line: Background terms. Solid line: Full model with
set FF1 of nucleon-Roper transition form factors.  Data from
Ref.~\protect\cite{Kitagaki:1986ct} (solid circles) and
Ref.~\protect\cite{Day:1984nf} (open squares).
\label{fig:5}}
\end{figure}

We present the results for the cross section for the $\nu n\to \mu^- p
\pi^+\pi^-$ channel in left panel of Fig.~\ref{fig:5} and for the
channel $\nu p\to \mu^- n \pi^+\pi^+$ in the right hand panel.  The
phase space for these results was restricted following a suggestion by
Adjei {\it et al.}\cite{Adjei:1980nj}.  We show our results for the
first channel with only background terms and with the full model
evaluated using the set FF1 of nucleon-Roper transition form
factors. Other sets give a similar result in this case. Even in this
kinematic region, the theoretical results including the resonance
contribution are lower than the experiment.  For the second channel
there are no contributions from the $N^*(1440)$ resonance.

\end{document}